\documentclass[final,5p,times,twocolumn]{elsarticle}

\usepackage{amssymb}
\usepackage{graphicx}

\journal{Vacuum}

\begin{document}
	
\begin{frontmatter}

\title{Energy deposition and formation of nanostructures in the interaction of highly charged xenon ions with gold nanolayers}

\author[a]{I. Stabrawa}
\author[a] {D. Bana\'{s} \corref{mycorrespondingauthor}}
\cortext[mycorrespondingauthor]{Corresponding author}
\ead{d.banas@ujk.edu.pl}
\author[a]{A. Kubala-Kuku\'{s}}
\author[a]{{\L}. Jab{\l}o\'{n}ski}
\author[a]{P. Jagodzi\'{n}ski}
\author[a]{D. Sobota}
\author[a]{K. Szary}
\author[a]{M. Pajek}
\author[b]{K. Skrzypiec}
\author[b]{E. Mendyk}
\author[c]{M. Borysiewicz}
\author[d]{M. D. Majki\'{c}}
\author[e]{N. N.  Nedeljkovi\'{c}}

\address[a]{Institute of Physics, Jan Kochanowski University, Uniwersytecka 7, 25-406 Kielce, Poland}
\address[b]{Departament of Chemistry, Maria Curie-Sk{\l}odowska University, Plac M. Curie-Sk{\l}odowskiej 3, 20-031 Lublin, Poland}
\address[c]{Institute of Electron Technology, aleja Lotników 32/46, 02-668 Warszawa, Poland}
\address[d]{Faculty of Technical Sciences, University of Pri\v{s}tina in Kosovska Mitrovica, Knjaza Milo\v{s}a 7, 38220  Kosovska Mitrovica, Serbia}
\address[e]{Faculty of Physics, University of Belgrade,  P.O. Box 368, 11001 Belgrade, Serbia}

\begin{abstract}
The effect of the deposition of kinetic energy and neutralization energy of slow highly charged xenon ions on the process of the nanostructures creation at the surface of gold nanolayers is investigated. The nanolayers of thickness of 100 nm were prepared by e-beam evaporation of gold on crystalline silicon Si(100) substrate. The samples were irradiated at the Kielce EBIS facility of the Jan Kochanowski University (Kielce, Poland), under high vacuum conditions. The irradiations were performed for constant kinetic energy 280 keV and different ions charge states (Xe\textsuperscript{q+}, q = 25, 30, 35, 36 and 40) and for constant charge state Xe\textsuperscript{35+} and different kinetic energies: 280 keV, 360 keV, 420 keV and 480 keV. The fluence of the ions  was on the level of 10$^{10}$ ions/cm$^{2}$. Before and after irradiation the nanolayer surfaces were investigated using the atomic force microscope.

As the result, well pronounced modifications of the nanolayer surfaces in the form of craters have been observed. A systematic analysis of the crater sizes (diameter on the surface and depth) allowed us to determine the influence of the deposited kinetic and the neutralization energy on the size of the obtained nanostructures. The results are theoretically interpreted within the micro-staircase model based on the quantum two-state vector model of the ionic Rydberg states population. The charge dependent ion-atom interaction potential inside the solid is used for the calculation of the nuclear stopping power. According to the  model the formation of the nanostructures is governed by the processes of the ionic neutralization in front of the surface  and the kinetic energy loss inside the solid. The interplay of these two types of processes in the surface structure creation is described by the critical velocity. Using the proposed theoretical model, the neutralization energy, deposited kinetic energy and critical velocities were calculated and compared qualitatively with the experimental results. The results are consistent (after normalization) with previous experimental data and molecular dynamics simulations for single ionized Xe and crystalline gold surface.

\end{abstract}

\end{frontmatter}

\section{Introduction}

Modification of metal, semiconductor and insulator surfaces by the ion irradiation is of great importance for developing new technologies for manufacturing a small functional electronics systems with nanometer dimensions, and has the potential to introduce novel nanostructures and material properties not achievable by any other material processing methods \cite{Barth2005}. Modification of materials by swift (high kinetic energy) heavy ion (SHI) irradiation is already used in many industrial processes, such as: the generation of nanopores in polymers \cite{Apel2003}, controlled drug delivery in biomedicine \cite{Rao2003}, precise band gaps modification \cite{Devaraju2010, Choudhury2013}, modification of high temperature superconductors \cite{Wiesner1994}, and others \cite{Spohr1998, Rizza2015}. It has also been demonstrated that with SHI beams regular patterns (usually in amorphic form \cite{Chan2007}) of lateral dimensions in the order of several tens of nanometers can be created.

One of the promising alternatives for creation of surface nanostructures is modification of surface by an impact of a single (i.e. each ion creates nanostructure) low-energy (slow) highly charged ions (HCI). The term slow HCI usually refers to impact velocities $v\ll$ 1 a.u., corresponding to 25 keV/amu (nuclear stopping power regime). HCI are characterized by an additional (to the kinetic energy) high potential  energy, resulting from the removal of many of the electrons from the neutral atom. For example, for Xe$^{50+}$ ion the potential energy is around 100 keV, i.e. 8400 times higher than that of a single charged xenon Xe\textsuperscript{+} ion. The neutralization energy of the HCI in the interaction with solid surface is also large for very slow ions (in keV energy range). As a consequence, the interaction of slow  single HCI with a surface is also governed by the potential (neutralization) energy of the ion \cite{Lake2011, Wilhelm2015, Majkic2021}. This energy is deposited on a small surface area along the first few nanometers below the target surface \cite{Aumayr2004}. Recent research on 2D materials shows \cite{Sch20}, that potential energy deposition of highly charged ion (Xe$^{38+}$) is limited to only up to two layers within multilayer MoS$_{2}$ (on graphene). For very low ionic velocities (down to $v= 0.03$ a.u.) the deposited potential energy (close to the ionic neutralization energy) can lead \cite{Facsko2009} to creation of various surface nanostructures, so far mainly observed on insulators such as alkali and alkaline earth halides, oxides and polymers, but also on highly oriented pyrolytic graphite (HOPG), sapphire and gold crystals, and silicon semiconductor \cite{Aumayr2011}. On the other hand, for moderate ionic velocities ($v \approx 0.25$ a.u.) both the neutralization and the deposited kinetic energy participate in the surface modification \cite{Lake2011, Majkic2021}.
Nanostructures created using HCI can have a form of hillocks, craters (called also pits) or caldera-like structures, with diameter of about 5-20 nm and a few nanometers vertical extension \cite{Facsko2009, Aumayr2011, Wilhelm2015}. It is known from experiments, that different parameters of ion beams, type of irradiated materials and processing conditions lead to different characteristic of the modifications obtained on a material surface, including defect production, sputtering of material and changes in material surface topology.

The recent studies of nanostructures formation on surfaces by HCI concentrate mainly on the basic characterization of nanostructures and fundamental understanding of the mechanisms responsible for the surface modifications \cite{Facsko2009, Lake2011, Majkic2021}. Moreover, most of the experimental observations were performed for insulator while for semiconductors (pure Si) and metals (Ti, Au) only single experiments were carried out, which due to the lack of the systematic studies did not allow for a detailed examination of the mechanism of nanostructures production on such surfaces \cite{Aumayr2011}. The reason for the small interest in this type of studies were the earlier experiments with swift heavy ions (SHI), which suggested that in the interaction of such ions with materials of high thermal conductivity, the production of nanostructures is unlikely due to the rapid outflow of energy from the area of impact. However, the results of experiments performed by Pomeroy et al. \cite{Pomeroy2007} and our recent results \cite{Stabrawa2017} showed that different nanostructures can be produced by slow single HCI also on metallic surfaces. Unfortunately in both of these experiments potential and kinetic energies of the ions were simultaneously changed, which made it difficult to separate their influence on the produced nanostructures.

Systematic experimental studies of the interaction mechanism are very important also from the theoretical point of view because there is still no unified picture of the nanostructure creation process. Up to now, the proposed theoretical models of the nanostructure production by slow single HCI, including Coulomb explosion \cite{Fleischer1965, Parilis1996, Parilis2001}, molecular dynamics simulations \cite{Nordlund2014, Khara2017}, inelastic thermal spike model \cite{Toulemonde1992, Dufour2017}, and plasma model \cite{Ritchie1982, Insepov2008}  describe the mechanism only in a qualitative manner and agree quantitatively only with the results of selected experiments, mainly for insulators. For the metallic surface  modifications, the  micro-staircase model of the HCI neutralization accompanied by the charge dependent model of the kinetic energy loss has been proposed \cite{Nedeljkovic2016, Majkic2021}.

The aim of the present study  is the systematic experimental and theoretical investigation of the mechanism of energy deposition and nanostructures creation in collisions of a single HCI with metallic surfaces. We consider the moderate ionic velocity region, characterized by the interplay of the neutralization  energy and the deposited kinetic energy. We performed the experiment with Xe\textsuperscript{q+} ions (q = 25, 30, 35, 36 and 40) impinging upon a gold nanolayer at kinetic energy 280-290 keV and with Xe$^{35+}$ ions  at kinetic energies: 280 keV, 360 keV, 420 keV and 480 keV. As the results, we obtained the well pronounced modification of the surface in the form of craters. In the present paper, the results are interpreted within the prediction of the micro-staircase model \cite{Nedeljkovic2016, Majkic2021} and molecular dynamics simulations for single ionized xenon hitting crystalline gold surface \cite{Bringa2001}. According to the micro-staircase model, simultaneously with the ion cascade neutralization above the surface, the neutralization energy deposits into the solid inducing the first destabilization of the target as a consequence of the high free electron density characteristic for conducting surfaces. Below the surface, the kinetic energy loss is governed by the elastic collisions between the ion (carrying the information about the ionic initial charge and velocity) and target atoms.

This article is organized as follows. In Section 2 we discuss the energy deposition process during the interaction of HCI with surface and current status of the experimental and theoretical studies for single and highly ionized xenon atoms  interacting with metallic (gold) surface. In this section we also introduce the micro-staircase model of the HCI-metal interaction. Section 3 is devoted to the present experiment. We characterize samples, describe the experimental conditions and the atomic force microscope (AFM) system used for the sample imaging. In Section 4 we present examples of the AFM images and extracted diameters and depths of the observed craters. In Section 5 we discuss the results and compare them with theoretical predictions and available experimental data for single ionized xenon \cite{Donelly1997}. The concluding remarks are given in Section 6.

\section{HCI - surface interaction}

\subsection{Overview of the HCI interaction with metallic surfaces}

Up to now, nanometer-sized structures produced by individual HCI impact on conductive surfaces were reported for a crystalline Au(111) by Pomeroy et al. \cite{Pomeroy2007}. In this experiment, the samples were irradiated with 200 keV Xe\textsuperscript{25+} and 350 keV Xe\textsuperscript{44+} ions, which have significantly different potential energies, 8 keV and 51 keV, respectively. After irradiation the samples were analyzed in situ with scanning tunneling microscope (STM). The STM images showed many different features on the gold surface, such as  isolated hexagons, hexagonal rings with craters in the center and hexagonal islands with pits, with the features density approximately equal to the ions fluence. It is worth to note  that  previous sputtering measurements \cite{Hayderer2001} with gold did not report a measurable increase in sputter yield with increasing of the HCI charge and thus probability for a nanostructure formation on gold was assumed to be negligible. Finally, Pomeroy et al. concluded that the primary formation mechanism of the features they observed on Au(111), is related to the kinetic energy (nuclear energy loss) and seems weakly dependent on the potential energy of the HCI, they emphasized the simultaneous change in the potential and kinetic energy of the ions used in the experiment, which complicated to isolate their contribution to the created nanostructures. Subsequent attempt to repeat Pomeroy et al. experiment using 440 keV Xe\textsuperscript{44+} ions has failed, probably due to too high surface roughness \cite{El-Said2007c}.

\begin{figure*} [!t]
	\centering
	\includegraphics[width=15cm]{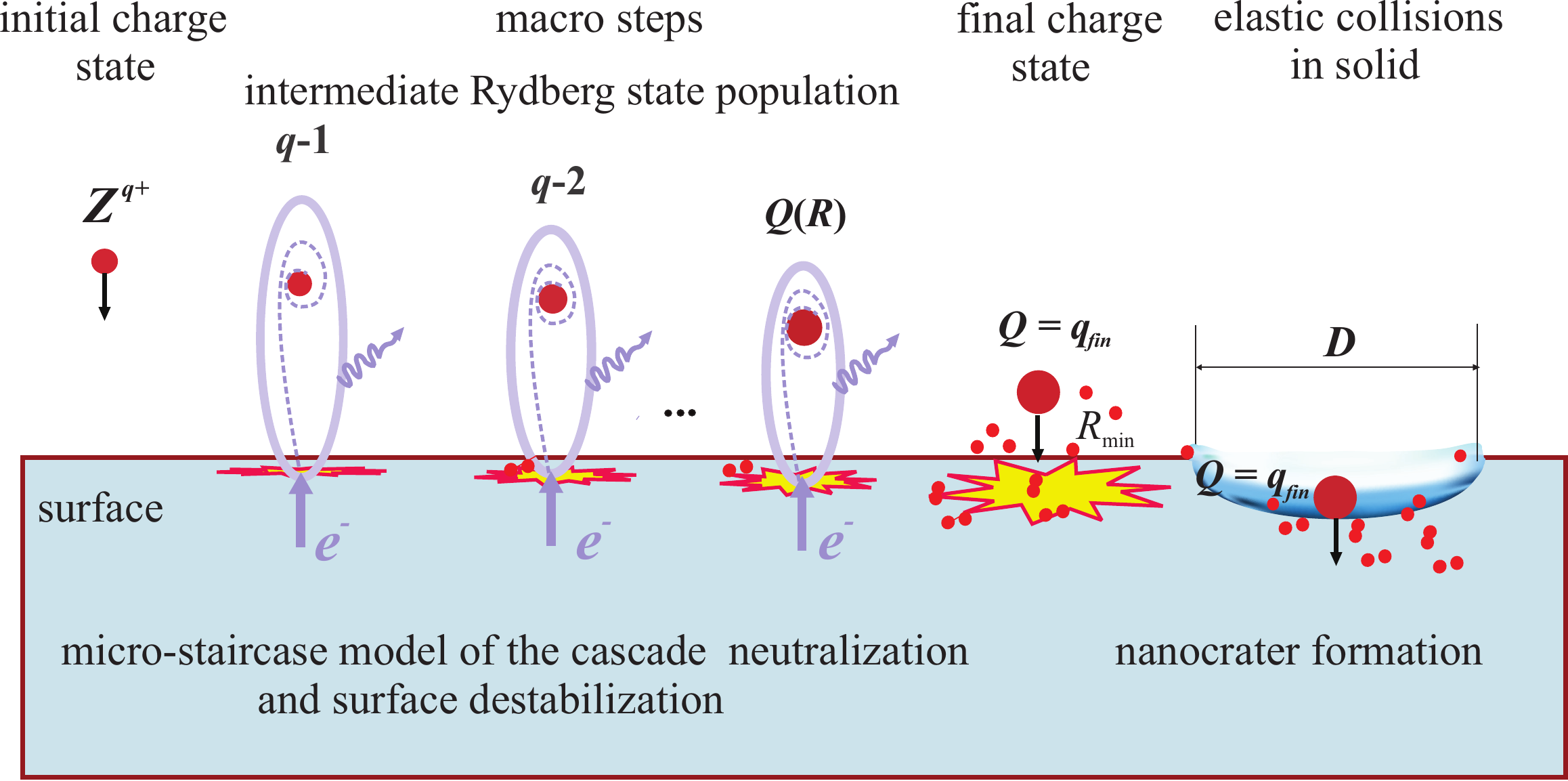}\\
	\caption{Schematic description of the micro-staircase model of the cascade neutralization with intermediate Rydberg state population followed by rapid deexcitation (both presented by dashed curves) and the nanocrater formation processes during the interaction of HCI with solid surface \cite{Majkic2021}.
		\label{fig:HCI_nanostructure2}
	}
\end{figure*}

A similar experiment, but at lower velocities, was also carried out by our group \cite{Stabrawa2017}. In this experiment, nanolayers of gold and titanium, were irradiated with low-energy (50-120 keV) highly charged xenon ions. The samples were prepared at Institute of Electronic Materials Technology, Warsaw, Poland, by sputtering of gold (50 nm) and titanium (75 nm) nanolayers on polished crystalline quartz SiO\textsubscript{2}(100) 4-inch diameter wafers, and titanium (25 nm - 75 nm) nanolayers on crystalline silicon Si (100) wafers. The samples were irradiated at the Kielce EBIS facility (Institute of Physics, Jan Kochanowski University, Kielce, Poland) \cite{Banas2015}. As a result of irradiation, we were able to create nanohillocks on both titanium and gold surfaces and perform statistical analysis of their heights and volumes using AFM images \cite{Stabrawa2017}. In this experiment the kinetic energy of the ions have been charge dependent (because of the ion source configuration) and thus it was difficult to extract separately the potential or the kinetic energy influence. 

A systematic analysis of the Xe\textsuperscript{q+} ion interaction with gold nanolayers at moderate velocities (craters formation) will be presented in Section 3. 

\subsection{Overview of the single charged ions interactions with metallic surfaces}

A many of scientists have discovered small craters on metallic surfaces bombarded with single ionized high-energy heavy ions which they attribute to the effect of spikes. The concept of thermal spikes resulting from single ion impacts was discussed for the first time in the  literature in the 1950s by researchers such as Brinkmann \cite{Brinkman1954}, Seeger \cite{Seeger1962}, and Seitz and Koehler \cite{Seitz1956, Koehler1961}. In particular, Merkle and J\"{a}ger used transmission electron microscopy (TEM) to examine Au surfaces irradiated with single ionized Bi and Au ions, in the energy range of 10-500 keV and discovered craters on the irradiated surfaces for the ion energies above 50 keV with fewer than 1\% of collisions causing the crater formation \cite{Merkle1981}. Average crater sizes were typically about 5 nm. Although the authors conclude that spike effects were responsible for the crater formation they attribute the effect mainly to sublimation of surface atoms from the surface \cite{Merkle1981}. Following this experiment, Birtcher and Donelly irradiated Au(110) films with Xe\textsuperscript{+} ions at energies of 50 keV, 200 keV or 400 keV. They found, using in situ TEM, that single xenon ion impacting on gold forms crater with size as large as 12 nm and that approximately 2-5\% of impinging ions produce craters \cite{Birtcher1996, Donelly1997}. Authors concluded that crater formation results from ion-induced sudden melting (and volume expansion) of the material associated with localized energy deposition (surface energy spikes) and explosive outflow of material from the hot molten core. The later experiments of Donelly and Birtcher on surfaces of Ag, In, and Pb led them to the same conclusions \cite{Donelly1999}.

The results of Donelly and Birtcher experiment \cite{Donelly1997} were examined using classical molecular-dynamics (MD) simulations by Bringa et al. \cite{Bringa2001}. They performed simulations of crater formation during 0.4-100 keV single charged Xe\textsuperscript{+} bombardment of Au target. The simulations confirmed that the craters are built by liquid flow of atoms from the interaction zone. They also found that energy density needed for crater production strongly depends on the heat spike lifetime and that for xenon energies higher than 50 keV cratering can results from lower energy densities due to long lifetime of the heat spike. MD simulated cratering probability was always higher than 50\% in the studied energy range \cite{Bringa2001}.

\subsection{Nanostructure formation on metallic surface: micro-staircase model}

Recently, in the article \cite{Majkic2021} we discussed the nanohillocks formation by the impact of Xe$^{q+}$ ions on titanium and gold nanolayers \cite{Stabrawa2017} using the micro-staircase model for the cascade neutralization based on the quantum two-state vector model (TVM).
The model takes into account both the ionic neutralization energy and the kinetic energy deposition inside the solid \cite{Nedeljkovic2016, Majkic2021}. The similar model can be used for the  analysis od the craters formation.

According to the model, the process of the cascade neutralization of the ion, $Q=q \to q-1 \to ...Q(R) \to ... q_{fin}$, is mainly localized in front of the surface (see Fig. \ref{fig:HCI_nanostructure2}). At ion-surface distance $R$, the electron is captured from the metal into the intermediate high-$n$ (Rydberg) state of the ion almost in a ground state. The population of each macro step consists of several micro-steps (population of the low-$l$ Rydberg states $n_Q$ with the probabilities $P_{n_Q}$). For example, considering the Xe$^{25+}$ ion impinging upon the metal surface at moderate velocity $v=0.25$ a.u. we have the following populate scheme \cite{Nedeljkovic2016}: at ion-surface distances $R$, in the range from $R=28 $ a.u. to $R=10$ a.u., the Rydberg states corresponding to $n=23$ to $n=15$  of the ion with charge $Q=q-1=24$ (core charge $25$) are populated with probabilities $P_{n_{25}}=0.01 \to 0.2$, $\sum P_{n_{25}}=1$. After the first macro-step is finished at $R=10$ a.u, the population of the ion of the charge $Q=q-2$ with core charge $24$ begins in the range from  $R=9 $ a.u. to $R=6$ a.u. The states $n=14$ to $n=12$ are populated with probabilities $P_{n_{24}}=0.3 \to 0.35$, $\sum P_{n_{24}}=1$, and so on. In Fig. \ref{fig:HCI_nanostructure2}) we present the final stages of the macrosteps $Q=q$, $Q=q-1$, ... $Q=q_{fin}$. At each macro step, the rapid deexcitation could be via radiative process and closer to the surface via Auger type processes with secondary electron emission in interplay with the described population process \cite{Nedeljkovic2016}. The  neutralization cascade  finishes when the HCI arrives into the interaction region at minimal ion-surface distance $R=R_{min}$ \cite{Majkic2021} with the final charge $q_{fin}$ ($R_{min}$ is the distance from the jellium edge \cite{Majkic2017}). The corresponding  neutralization energy $W^{(q,nMV)}$ within  the nanolayer-metal-vacuum (nMV) system is deposited into the first nanometers of the surface \cite{Hatas1999} very fast (few fs for metal targets \cite{Wilhelm2015, Pomeroy2007}), increasing the energy density in the impact region \cite{Lemell2007} and inducing the destabilization of the surface \cite{Majkic2021}. 

The neutralization energy $W^{(q,nMV)}$ is defined as a difference between the potential energy $E_p \equiv W_{q,pot}$ (which describes the state of the ion before the beginning of the neutralization) and the potential energy in front of the solid surface $W_{q_{fin}^{nMV},pot}$ \cite{Majkic2017, Majkic2019, Majkic2021}:

\begin{equation}\label{Neu}
	W^{(q,nMV)}=W_{q,pot}-W_{q_{fin}^{nMV},pot}.
\end{equation}

The energy $W^{(q,nMV)}$ can be calculated using the results valid for the metal-vacuum MV-system \cite{Majkic2021}, which is supported by the experimental fact that the lattice structure of the  nanolayer  is very similar to the bulk material for the  layers thickness \cite{Siegel2011} considered in the present article.

Although many elementary charge exchange processes are possible at solid surface, we assume that the ion with $Q=q_{fin}$ penetrates the surface. Within the framework of the model, we also assume that neutralization inside the solid in the process of the nanostructure formation can be neglected (has negligible influence on the total deposited energy). The last assumption is based on the fact that the nanocraters are formed in the narrow region of the depth $\Delta x$ smaller compared to the penetration depth necessary for the ionic charge to be significantly changed \cite{Majkic2021}. Therefore, we use $Q\approx q_{fin}$ as the ionic charge for the analysis of the ionic motion and the corresponding processes inside the target (see Fig.2). For more accurate description of the overall neutralization process, the analysis of the neutralization below the surface can be added to our model.

Below the surface, the ions constantly lose their kinetic energy due to the elastic collisions with the target nuclei (nuclear stopping power $dE_n/dx$) \cite{Wilhelm2015,Moler2017,Nordlund2010} and the inelastic interaction with the target electrons (electronic stopping power $dE_e/dx$). Simultaneously, the damage of the local atomic structure and the surface modification due to the ionic kinetic energy loss take place. On the overall ionic trajectory the ionic kinetic energy is deposited into the solid. However, analyzing the experimentally obtained surface nanostructures,  relevant is only the near surface region of the length  $\Delta x$  \cite{Majkic2021}, so that
\begin{equation}\label{Ek}
	E_{k,dep}=(dE/dx)\cdot \Delta x.
\end{equation}
\noindent
For low to moderate ionic velocities electronic stopping power can be neglected, i.e.,  $dE/dx=dE_n/dx =N S_n$, where $N$ is the atomic density of the target.  The corresponding nuclear stopping cross section $S_n$ we calculate using the classical scattering theory with charge dependent ion-target atom interaction potential \cite{Wilhelm2016II,Majkic2021}:
\begin{equation}\label{V}
	V_{int}(r)=\frac{(Z_1-Q)Z_2}{r}\varphi(\frac{r}{a_u})+\frac{Q Z_2}{r}\varphi (\frac{r}{a_s}),
\end{equation}
\noindent
where $Z_1$ and $Z_2$  are the nuclear charges of the projectile and the target atom, respectively. Other quantities are explicitly given in \cite{Majkic2021}.
We note that, the charge dependence of the energy loss has been firstly theoretically introduced by Biersack \cite{Biersack}. Further, the charge dependent kinetic energy transfer for HCI interacting with C nanomembrane and C foil target was elaborated in \cite{Lake2015}. The time-dependent interatomic potential energy was used to get the more accurate model for the calculation of the kinetic energy loss in \cite{Wil19}.

\section{Experiment}

The studies presented in this paper are continuation of our research \cite{Stabrawa2017, Stabrawa2020} related to the nanostructure formation in interactions of highly charged xenon ions with metallic surfaces. The aim of the experiments carried out for the purposes of current work is to separate, as far as it is possible, the influence of kinetic and potential energies of the HCI xenon ions on the produced surface nanostructures. In the measurements we use gold nanolayers of the thickness of 100 nm deposited on Si (110) wafers. The structure and properties of such nanolayers were expected to be similar to the bulk metal, but with possible lower density due to nanolayer structure \cite{Siegel2011} and an influence of the substrate cannot be completely excluded \cite{Lake2011}.

\subsection{Samples}
The 100 nm Au nanolayers used in this experiment were prepared at Institute of Electron Technology (Warsaw, Poland) using high vacuum (13$\cdot$10$^{-8}$ hPa) e-beam evaporation VST TFDS-462U deposition system. The metallic nanolayers were evaporated on Topsil (Warsaw, Poland) Si (110) polished prime wafers type N 4-inch diameter. The thickness of the silicon wafer was 0.635 mm $\pm$ 0.015 mm. The deposition rate was 0.4 nm/s. The thickness of the gold was set and controlled by a crystal oscillator. Just after preparing, the wafers were cut into rectangles of dimensions of 0.5 cm x 1 cm. The roughness of the samples surface was checked by AFM technique. Root-mean-squared (RMS) roughness determined by the AFM technique (UMCS, Lublin, Poland)  for a few randomly selected (1 $\mu$m x 1 $\mu$m) areas of the gold nanolayers using NanoScope Analysis ver.1.40 (Veeco, USA) program were on the level of $0.45\pm 0.1$ nanometers. The crystalline structure of the substrate was confirmed based on the measurements carried out using the XRD technique. Using the GIXRD technique, it was determined that the 100 nm gold nanolayers have a polycrystalline structure homogeneous in depth. Additionally using the XRR technique, the thickness and density of the nanolayers were measured. Thickness turned to be consistent with the declared one (100 nm), but the density was slightly lower (but within the uncertainties) than the bulk density. The XRR, XRD and GIXRD measurements were performed with X'Pert Pro MPD reflectometer/diffractometer (for details see \cite{Stabrawa2016, Stabrawa2018}), placed at Institute of Physics of UJK (Kielce, Poland). The 100 nm Au nanolayers were irradiated at the Kielce EBIS facility of the Jan Kochanowski University (Kielce, Poland) \cite{Banas2015}, under high vacuum conditions. After irradiation the samples were again checked with AFM technique (UMCS, Lublin).

\begin{figure*}
	\centering
	\includegraphics[width=17cm]{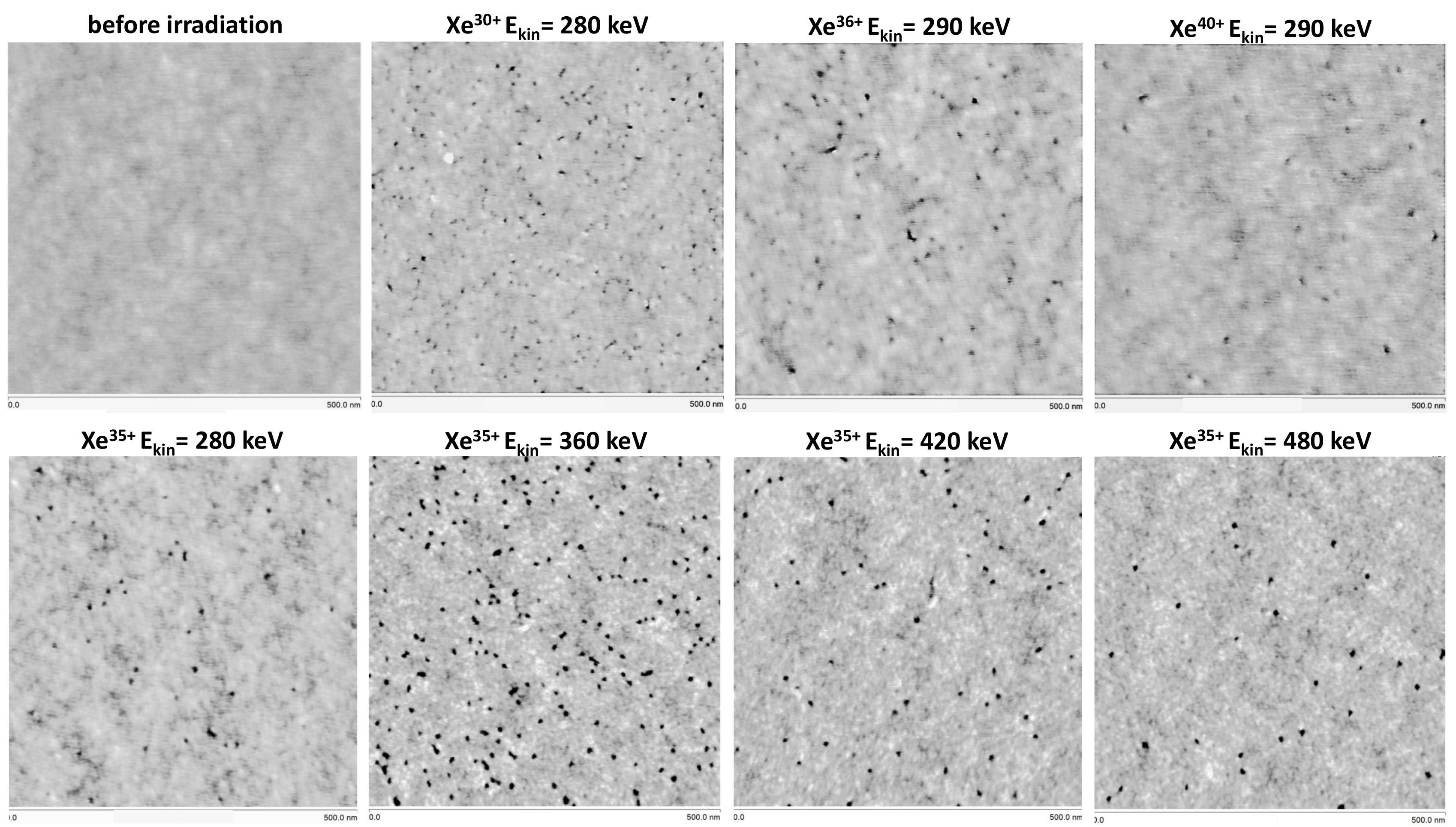}\\
	\caption{Topographic AFM 3D images of Au 100 nm nanolayer deposited on Si surface before and after irradiation with HCI Xe\textsuperscript{q+}. Top row: images of the nanolayers before irradiation (left panel) and after irradiation with 280-290 keV Xe\textsuperscript{30+}, Xe\textsuperscript{35+}, Xe\textsuperscript{40+}. Bottom row: images of the nanolayers after irradiation with Xe\textsuperscript{35+} of different kinetic energies. The images were analysed using the software NanoScope Analysis ver. 1.40 (Veeco, USA).
		\label{fig:Gold_clean_irradiated}
	}
\end{figure*}

\subsection{Kielce EBIS facility} \label{subsection:hci}
The Kielce EBIS facility, built by the Dreebit (Dresden, Germany), is equipped with electron beam ion trap (EBIS-A) \cite{Zschomack2006}. The source supplies a wide range of slow HCI from bare ions of light elements to Ne-like and Ar-like ions of high-Z elements. The maximum electron energy and current available for ionization of the trapped ions are equal 25 keV and 200 mA, respectively. The ions produced in the EBIS-A source can be extracted both in a pulse mode (pulse width from 2 $\mu$s up to 40 $\mu$s) and leaky mode (DC mode) by applying an acceleration voltage up to 30 kV. Highly charged ions extracted from the EBIS-A ion source are guided by ion beam optical elements (einzel lens and  X-Y deflectors) of the first straight section of the facility to the double focusing analyzing magnet separating the ions according to their mass to charge ratio. The first section of the beamline includes a quadrupole section with pressure gauge, 4-jaw-slit collimation system and a Faraday cup. The ions separated in the analyzing magnet are directed to the second straight section of the EBIS-A facility. In this section, a pressure gauge, X-Y deflectors, a Faraday cup and an einzel lens are mounted. Finally, the highly charged ions collide with a sample mounted on a 5-axis universal manipulator placed in the experimental chamber. The manipulator allows for x, y, z linear movements, polar and azimuthal rotations of a sample and variation of its temperature in the range of 100-1000 K. The beamline can be biased with positive or negative high voltage allowing ion acceleration or deceleration. For current facility configuration the ion energies can be set from 2.5 keV x q up to 30 keV x q, with q denoting the ion charge state. All components of the EBIS-A facility fulfill the UHV standards and after baking of the system at 150$^\circ C$ the pressure is in the few 10$^{-10}$ mbar range (in the beamline). One of the unique features of the EBIS facility is the ability to prepare, irradiate by highly charged ions and characterize the studied samples in the UHV conditions.

\subsection{Measurements}
In the measurements isotopically pure highly charged Xe\textsuperscript{q+} ions were extracted from the EBIS-A and, after selecting given ion charge state in the dipole magnet, were used to irradiate the nanolayers. The ion beam current was measured with a movable Faraday cup mounted in front of the sample. The spot radius of the ion beam on the sample was around 1.5 mm $\pm$ 15\% as it was determined by moving the Faraday cup across the ion beam (from the beam profile). The ion fluence was estimated on the level 10$^{10}$ ions/cm$^{2}$ (with uncertainty of the 10-15\%). The samples were first placed in a loading chamber pumped to about $10^{-7}$ mbar, and then transmitted to the experimental chamber. The vacuum in the experimental chamber was around $(2-5) \times 10^{-8}$ mbar. After irradiation, the sample was transferred back to the loading chamber and was stored there until it was removed for atomic force microscopy investigations, which were performed in the air. The measurements were performed for two configuration: constant kinetic energy of the ions equal to ~280-290 keV and different charge states of the xenon ions Xe\textsuperscript{q+}, where q = 25, 30, 35, 36, 40 and constant charge state (Xe\textsuperscript{35+}) of the ions and different kinetic energies 280 keV, 360 keV, 420 keV and 480 keV.

\subsection{AFM system}
The topographic modifications of the samples surface induced by Xe\textsuperscript{q+} ions were investigated using atomic force microscopy in the Analytical Laboratory of Faculty of Chemistry, UMCS, Lublin, Poland. AFM measurements of the studied samples were performed using Multimode 8 (Bruker) AFM equipped with NanoScope software (Bruker-Veeco, USA). The AFM was operated in SCANASYST-HR fast scanning mode using SCANASYST-AIR-HR probe (Silicon Tip on Nitride Lever) (Bruker) with the cantilever of force constant k = 0.4 N/m. The lateral and vertical resolutions were 4 nm and 0.1 nm for the 1 $\mu$m x 1 $\mu m$, and 2 nm and 0.1 nm for the 500 nm x 500 nm images. The obtained images were analyzed with Nanoscope Analysis ver. 1.40 software (Veeco, USA).

\section{Results}

\subsection{AFM images}
The AFM  images of the nanolayers before irradiation (left panel) and after irradiation with 280-290 keV Xe\textsuperscript{30+}, Xe\textsuperscript{36+}, Xe\textsuperscript{40+}, and Xe\textsuperscript{35+} of different kinetic energies are presented in the Fig. \ref{fig:Gold_clean_irradiated}. The images were analyzed using the  NanoScope Analysis software. The size of presented area is 500 nm x 500 nm. In the images of the irradiated samples, we can clearly see the modifications caused by the ion impact. We would like to stress here, that such excellent images of metallic surface modification caused by HCI impact, to our knowledge, have never been registered. The same modifications were observed for all irradiated samples, with surface density of the nanostructures approximately equal to the ion fluence, i.e. one nanostructure per one HCl ion impact. Analogous efficiency of the nanostructure creation was observed by Pomeroy \cite{Pomeroy2007}.

\begin{figure} [b!]
	\centering
	\includegraphics[width=9cm]{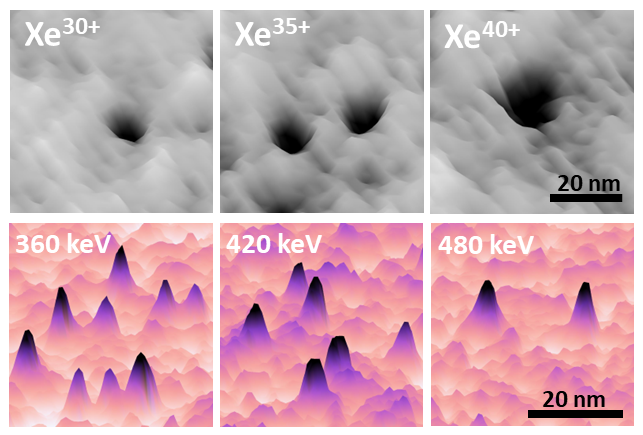}\\
	\caption{Upper panel: examples of the 3D AFM images of the nanostructures on Au 100 nm/Si nanolayer surface irradiated by 280-290 keV Xe\textsuperscript{30+}, Xe\textsuperscript{35+} and Xe\textsuperscript{40+}. Lower panel: upside-down 3D AFM images of the nanostructures on Au 100 nm/Si nanolayer surface irradiated by 360, 420 and 480 keV Xe\textsuperscript{35+}. The images were analysed using the software NanoScope Analysis ver. 1.40 (Veeco, USA).
		\label{fig:Gold_nanostructures}
	}
\end{figure}

The measured modifications have the form of craters, which is confirmed by enlarged AFM 3D images of the individual nanostructures which are presented in the Fig. \ref{fig:Gold_nanostructures}. In the upper panel the example of the 3D AFM images of the nanostructures on Au 100 nm/Si nanolayer surface irradiated by 280-290 keV Xe\textsuperscript{30+}, Xe\textsuperscript{35+}, Xe\textsuperscript{40+} are presented. All observed structures had a similar crater-like shape, i.e. a cavity, sometimes with a ring around it (check the middle image). Merkle and Jager \cite{Merkle1981} and Bringa et al. \cite{Bringa2001} postulate that these rings around the cavity arise from the sputtering (or rather an outflow) of the original atoms being at the place of the structure formation. This was confirmed by MD simulations presented in the article of Bringa et al. \cite{Bringa2001}. Similar shape of the crater formed on a Si(100) surface by bombardment of a Xe$^{44+}$ HCI was also observed in the simulations performed by Insepov et al. \cite{Insepov2008} using plasma model of space charge neutralization based on impact ionization of semiconductors at high electric fields. In the lower panel of the Fig. \ref{fig:Gold_nanostructures} upside-down 3D AFM images of the nanostructures on Au 100 nm/Si nanolayer surface irradiated by 360, 420 and 480 keV Xe\textsuperscript{35+} are presented to confirm crater like shape of the nanostructures.

\subsection{Analysis of the AFM images}
\begin{figure}
	\centering
	\includegraphics[width=8cm]{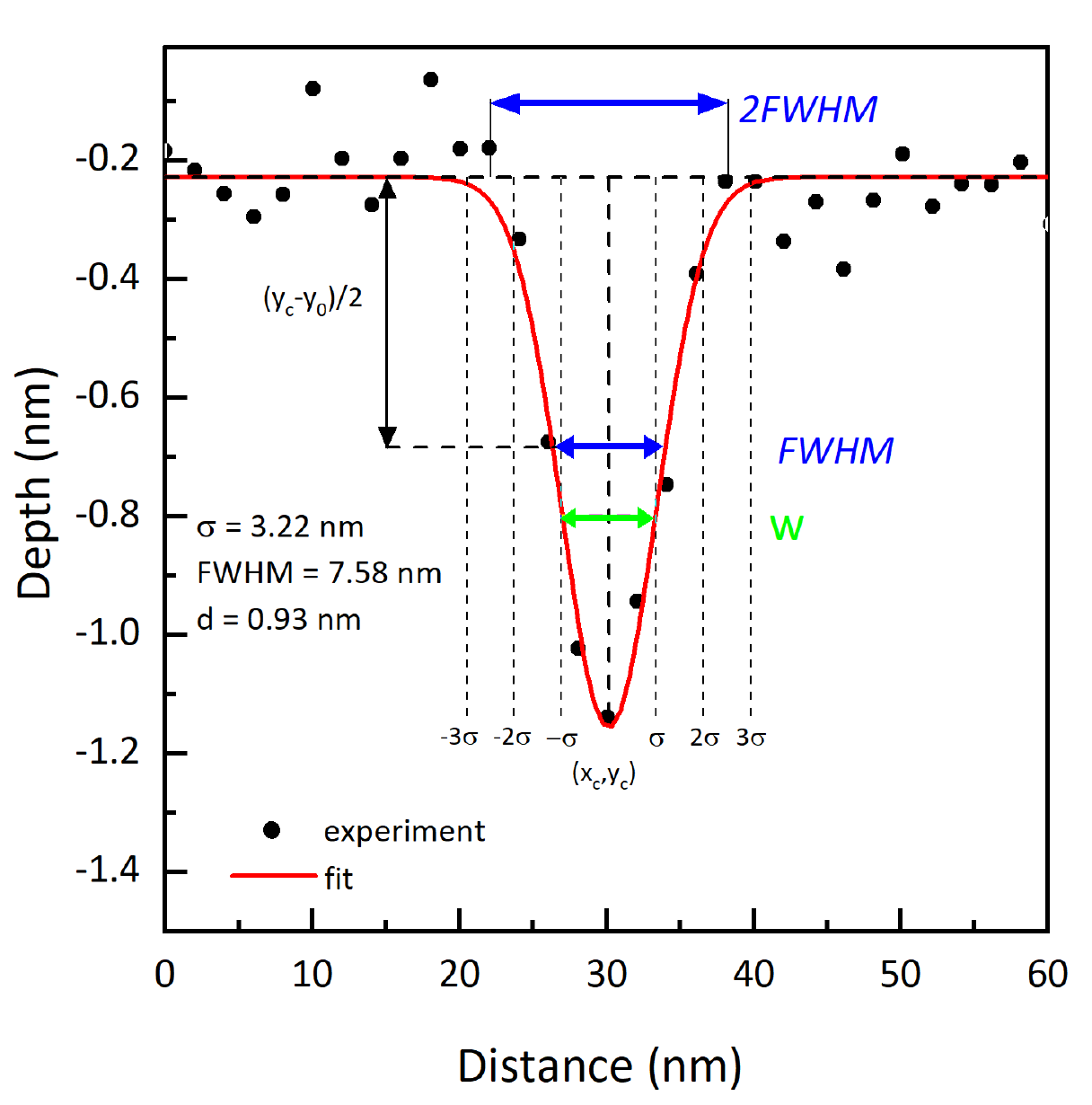}\\
	\caption{Example of the individual crater profile (black points) with fitted Gaussian curve (solid line). The sigma $\sigma$ (standard deviation), FWHM (full width in the half of maximum) and crater depth d quantities are Gaussian distribution parameters. The crater diameter on the surface is assumed as 2FWHM.
	\label{fig:Crater_profile}
	}
\end{figure}

In many surface studies, a common data analysis strategy is to correlate the mean size of nanostructures (parameters like: diameter, depth, volume) with different ion parameters, e.g. kinetic energy (ionic velocity) and potential energy (ionic charge state), nuclear and electronic stopping powers, etc. Following this strategy we have performed  size analysis of the observed nanostructures. For this purpose, all observed images were first carefully checked and optimized with NanoScope Analysis software and, after extracting of the data from the AFM images (using STEP function of the software), analysed with Origin Pro data analysis software. From the individual profile of the craters we have extracted their size parameters, including the diameter on the surface. All unambiguously identified structures on the surface of the samples were analyzed independently. Example of the individual crater profile (black points) is presented in the Figure \ref{fig:Crater_profile}.

The profiles were fitted by Gaussian curve (solid line), which reflected very well the shape of the crater. At this point, we note that the needle used for the AFM analysis of all samples had a tip curvature radius r = 2 nm, and the structures after HCI modification were characterized by diameters of about 10-25 nm, so an incorrect tip contact was considered unlikely, especially in the diameters of nanostructures on the sample surface. The crater diameter on the surface was defined as double FWHM (2 $\times$ FWHM). In the example presented in the Figure \ref{fig:Crater_profile}, the crater diameter at surface was fitted as 15.16 nm, while its depth as 0.93 nm. An alternative way to determine the diameter is to take values of four standard deviations (4$ \times \sigma$). In presented examples, it gives 4$\sigma$ = 12.88 nm. In general, it was observed that crater diameter defined as 2 $\times$ FWHM was about 10-15$\%$ higher than 4$ \times \sigma$ quantity.

For the Au nanolayer irradiated by Xe ions in given charge state, for each sample around 50 to 150 craters were analyzed in the way described above. Finally, the mean values of the crater depth and crater diameters for each irradiated Au nanolayer were calculated. Based on the statistical analysis of the crater profiles, the dependence of the crater depth and crater diameter on the Xe ions potential and kinetic energy were studied. In the case of the crater depth no dependence on the ions potential energy was observed. The crater depth was on the constant level of about 0.9 nm $\pm$ 0.15 nm. The linear fit to the data, including uncertainties defined by the standard deviation of mean value, gave a very week dependence (the slope is equal to 0.001 nm/keV).

\begin{figure} [b!]
	\centering
	\includegraphics[width=8cm]{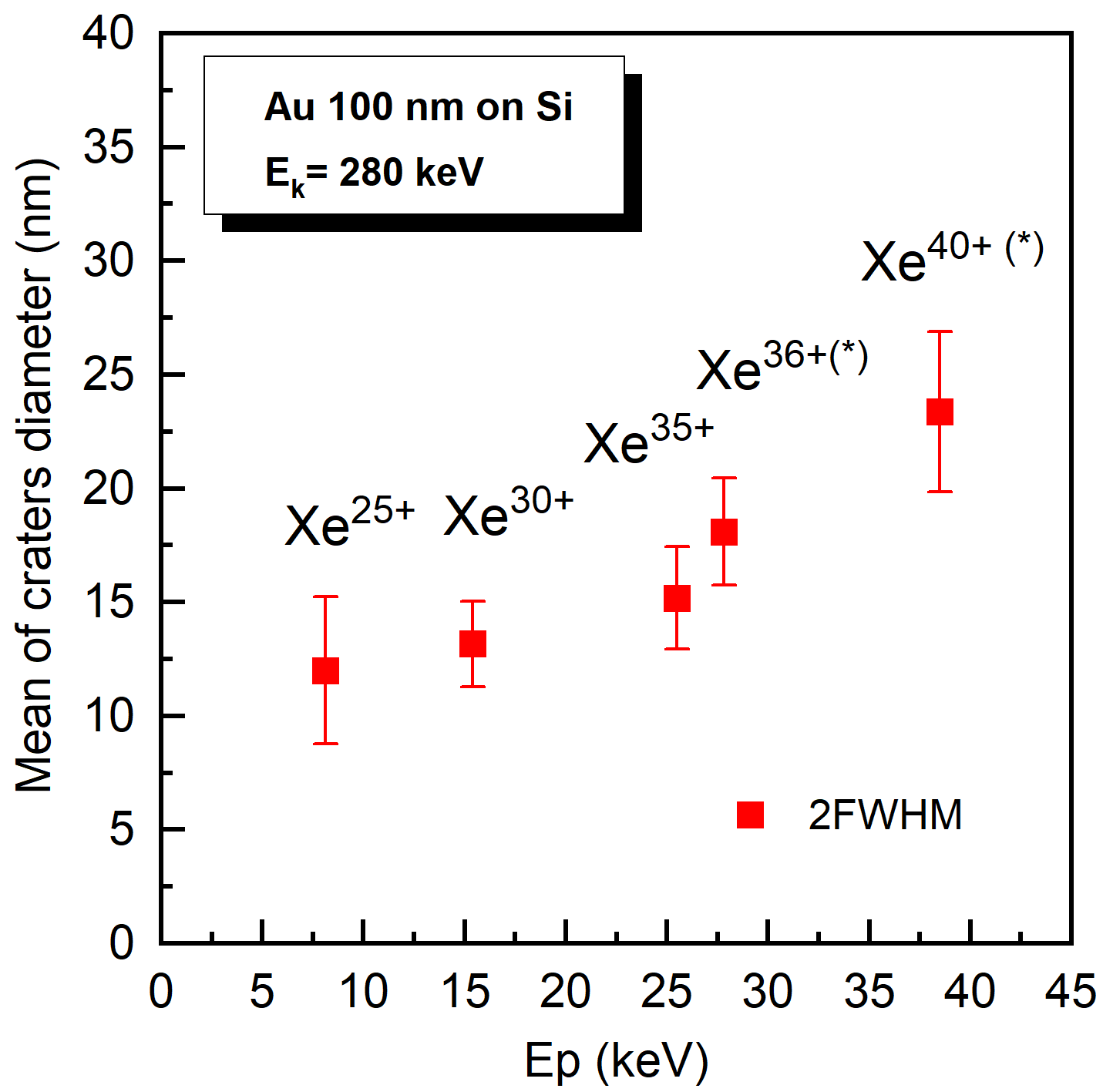}\\
	\caption{Dependence of the craters diameter on the Xe\textsuperscript{q+} ions potential energy. The Xe ions charge states marked by (*) denoted a slightly different kinetic energy (290 keV).
		\label{fig:Gold_potential_energy}
	}
\end{figure}

The obtained mean values of craters diameter in function of the potential energy of the Xe\textsuperscript{q+} ions are plotted in the Fig. \ref{fig:Gold_potential_energy}. The uncertainties marked for experimental points were calculated as the sum of the mean value standard deviation and 10$\%$ of the mean value (compensation of the difference between 2FWHM and 4$\sigma$ quantities within uncertainty). The Xe ions charge states marked by (*) denoted a slightly different kinetic energy (290 keV), caused by difficulties in setting a given charge state and kinetic energy.
As one can see from the figure we have observed clear influence of the ionic charge state (expressed via initial potential energy) on the nanocrater diameter. For the lowest ion charge state (25+) the mean nanocrater diameter is 12.0 nm, next this parameter systematically grows, reaching for the highest charge state the value 23.4 nm. 

\begin{figure} [t!]
	\centering
	\includegraphics[width=8cm]{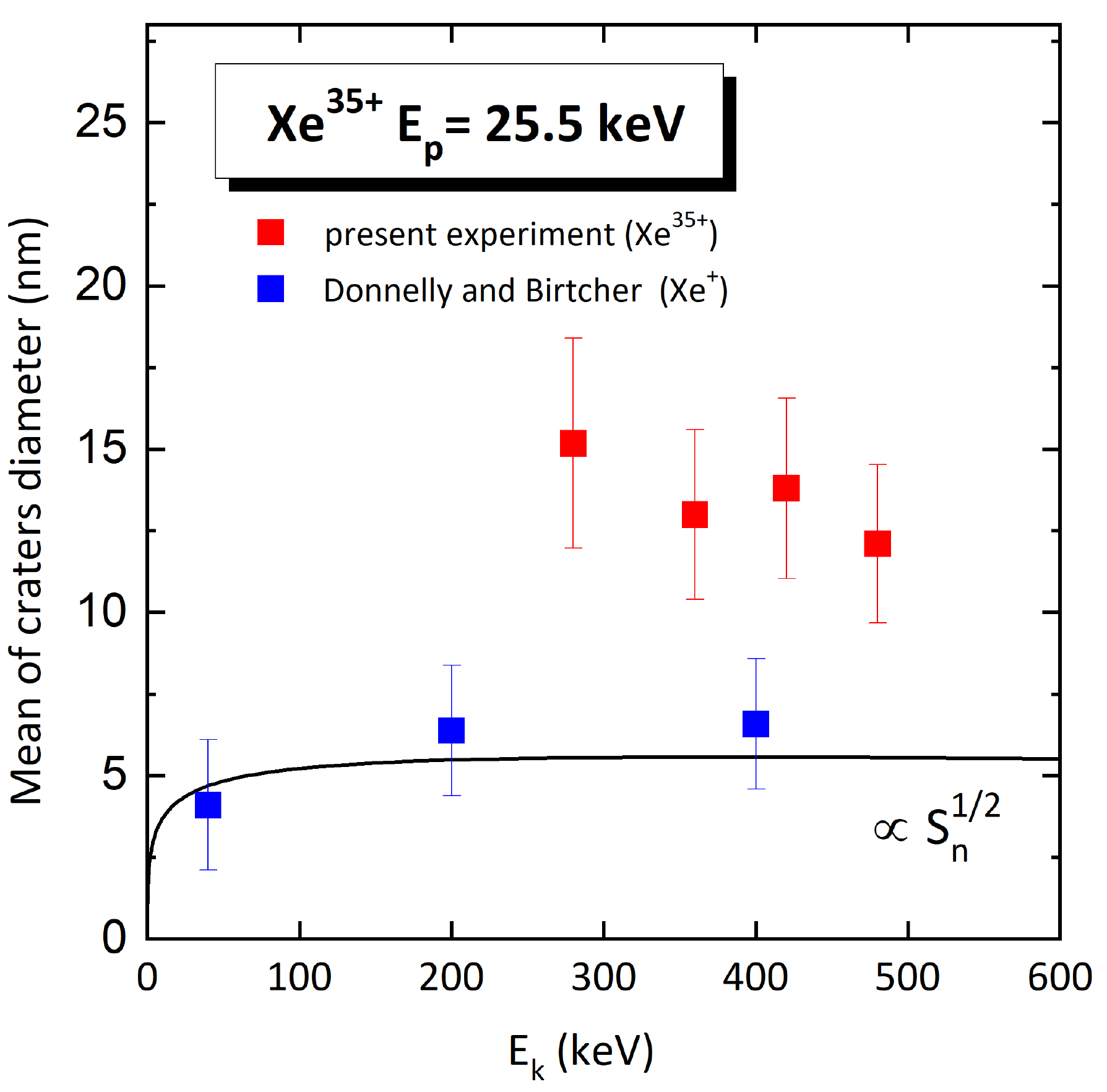}\\
	\caption{Dependence of the nanocrater diameter created by the Xe$^{35+}$ ions impinging on Au surface on the kinetic energy (this experiment). For comparison, the results of Donnelly and Birtcher experiment \cite{Birtcher1996} for Xe$^{+}$ ions are also presented. In the figure, also the nuclear stopping power S$^{1/2}$ (solid line) is presented.
	\label{fig:Gold_kinetic_energy}
	}
\end{figure}

The results of the study of the crater diameter in function of the ions kinetic energy are shown in the Figure \ref{fig:Gold_kinetic_energy} for Xe$^{35+}$. The nanocrater diameter is in the range 13-15 nm. The linear function fitted to the experimental points showed a weak alteration of the dependence. In the Figure  \ref{fig:Gold_kinetic_energy} the results of Donnelly and Birtcher experiment \cite{Birtcher1996} for Xe$^{+}$ ions are also presented which confirm small dependence of the created nanocrater diameter on the ions kinetic energy in the considered energy range. On the other hand, the nanocrater diameters for HCI xenon ions are much higher than for single ionized xenon.

\section{Discussion}

\subsection{Theoretical model of the crater formation}

In order to interpret the present experiments performed with Xe\textsuperscript{q+} ions of initial charges q = 25, 30, 35, 36 and  40 in the interaction with a gold 100 nm nanolayer deposited on Si (110) wafers (nMV-system) at velocity $v=0.29$ a.u., as well as to examine the velocity dependence studied in the case of Xe$^{35+}$ for $v=0.29, 0.33, 0.36$ and 0.38 a.u. we use the micro-staircase model.

The formation of nanocraters we discuss from the standpoint of the energy dissipation  into the  surface, which consists both of the neutralization energy and the deposited kinetic energy \cite{Majkic2017,Majkic2019,Majkic2021}. For velocities characteristic for the crater formation the neutralization is incomplete so that the corresponding neutralization energy represents only a part of the ionic initial potential energy. The remaining potential energy $E_p-W^{(q,nMV)}$ contributes to the charge dependent potential interaction (Eq.(\ref{V})) between the ion and the target atoms and thus it is converted into kinetic energy of the target atoms (stopping power calculated in micro-staircase model is charge dependent).
We calculate the neutralization energy according to Eq. (\ref{Neu}) for $W^{(q,nMV)}=W^{(q,MV)} $; taking into account that  the neutralization energy is weakly dependent on the solid work function $\phi$, we consider the neutralization energy for $\phi =5$ eV  (work function of Au is 5.47 eV). For the calculation of the kinetic energy loss we employ Eq. (\ref{Ek}) for the active interaction length $\Delta x \approx 5\bar{c} \approx 38.5$ a.u., where $\bar{c}$ is the mean lattice constant for Au-target; we note that the crater depth $d_{max}\approx 1 \textrm{nm} = 18.9$ a.u. To define the ion-atom interaction in solid we use the charge of the projectile $Q=q_{fin}(q,v)$ obtained in \cite{Majkic2021}.
\begin{figure}[!bt]
	\centering
	\includegraphics[width=8.2cm]{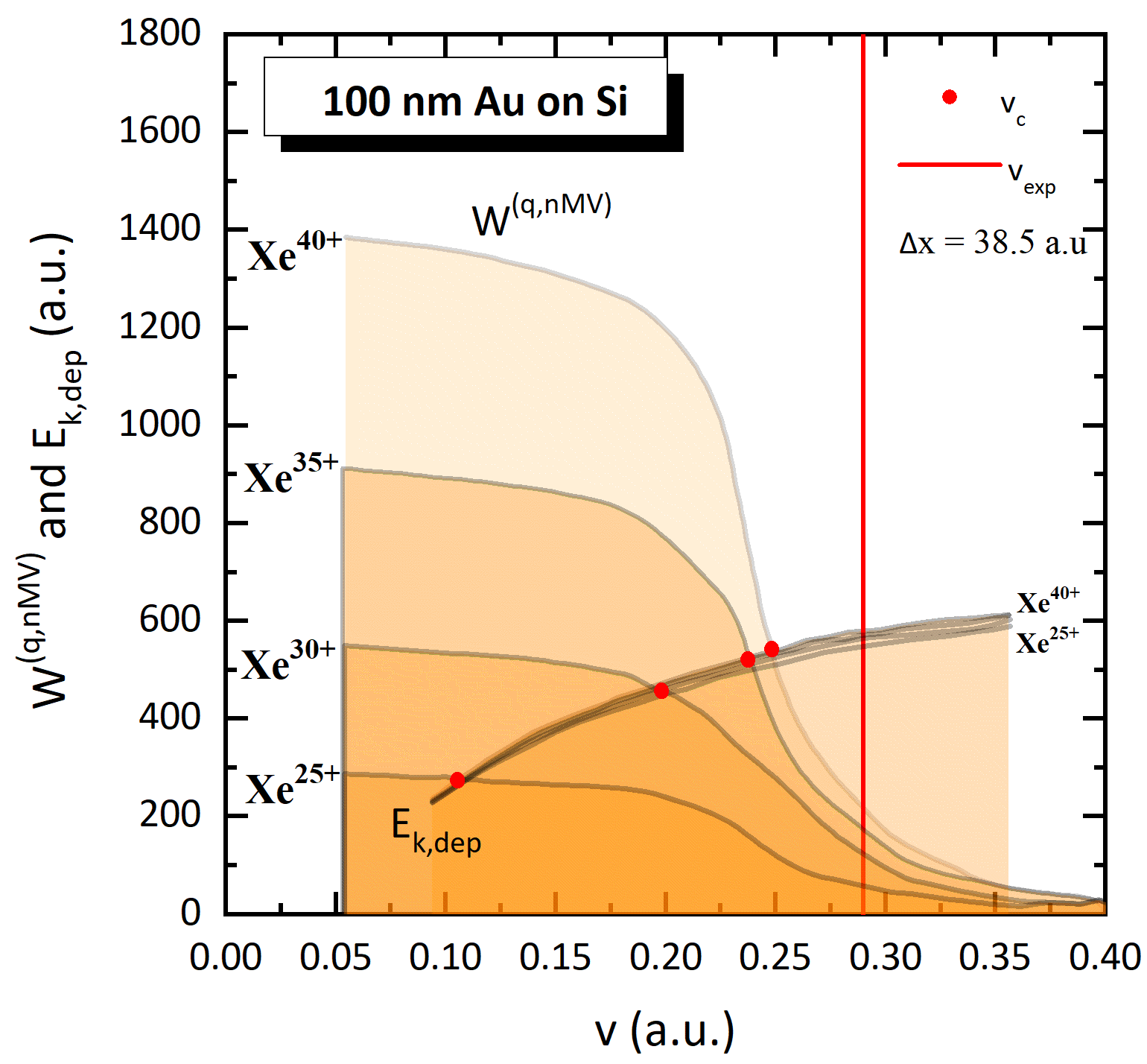}\\
	\includegraphics[width=8cm]{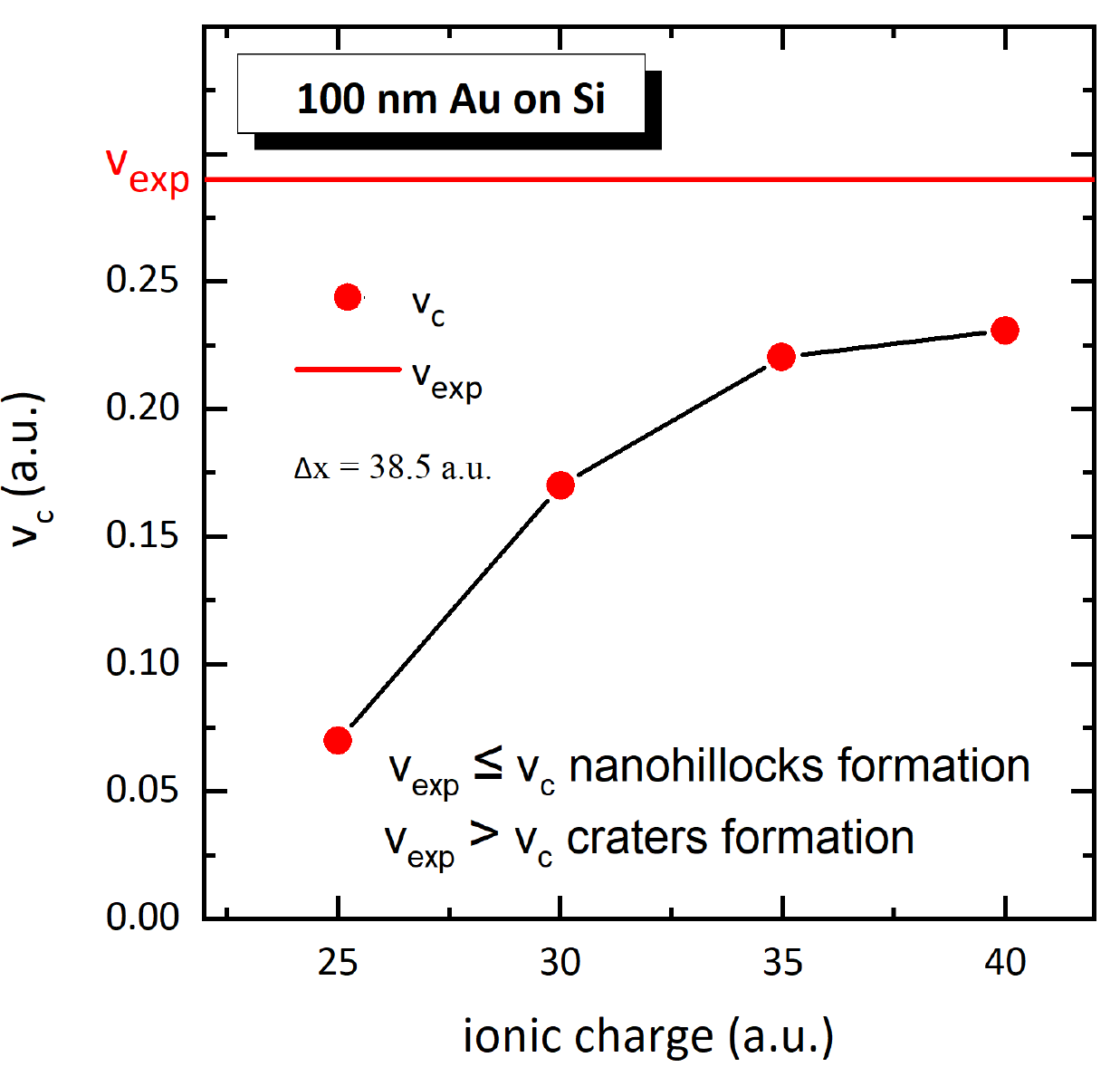}\\
	\caption{Upper panel: neutralization energy $W^{(q,nMV)}$ and  deposited kinetic energy $E_{k,dep}$ versus ionic velocity $v$ for Xe$^{q+}$ ions, $q=25,30,35$ and 40, impinging on the Au nanolayers (formation of the crater in the present  experiment). Lower panel: the critical ionic velocity $v_c$ versus initial ionic charge $q$.}
	\label{fig:figure8}
\end{figure}

In Fig. \ref{fig:figure8}, at upper panel, we present the neutralization energy $W^{(q,nMV)}$ and the  deposited  kinetic energy $E_{k,dep}$ relevant for the surface nanocrater creation by the impact of Xe$^{q+}$ ions with core charges $q=25,30,35$ and 40 on 100 nm Au nanolayer on Si (110) wafers as a function of the ionic velocity $v$. The neutralization energy $W^{(q,nMV)}$ decreases with increasing of the ionic velocity $v$; on the other hand, the deposited kinetic energy  $E_{k,dep}$ increases with increasing  of  $v$. The results indicate the interplay of these two energies in the process of the surface nanocrater formation.
That is, we define \cite{Majkic2021} the critical velocity $v_c$ by the relation:

\begin{equation}\label{v_c}
W^{(q)}(v_c)=E_{k,dep}(v_c).
\end{equation}

For velocities $v \ll v_c$ (very low ionic velocities)  dominant role in the energy participation in the solid has the neutralization energy $W^{(q,nMV)}$, while for $v \gg v_c$  (swift heavy ions) the deposited kinetic energy $E_{k,dep}$ completely determines the process of the nanostructure formation \cite{Majkic2021}. The quantity $v_c$ we present in Fig. \ref{fig:figure8} at lower panel as a function of the initial ionic charge $q$. The values of the critical velocities are also given in Table \ref{tab:table1}.

\begin{table}
	\begin{center}
		\caption{\label{tab:table1}
			Critical velocities $v_c$  in the case of the  surface nanocrater formation
			in the nMV-system by the impact of Xe$^{q+}$ ions. }
		\begin{tabular}{cccccccc}
			\hline
			\multicolumn{6}{c}{100 nm Au nanolayer} \\
			\hline
			$q$      &25  &30   &35  &40   & \\
			\hline
			$E _{k}$  (keV)   & 280  & 280   & 280  & 280  & \\
			\hline
			$v_{exp}$ (a.u.)   & 0.29  & 0.29   & 0.29  & 0.29 & \\
			\hline
			$v_c $ (a.u.)  & 0.07   & 0.17   & 0.22  & 0.23 & \\
			\hline
		\end{tabular}
	\end{center}
\end{table}

For all considered ionic charges the critical velocities $v_c $ are lower compared to the experimental value  $v_{exp}=0.29$ a.u. ($E_{k}=Mv_{exp}^2/2, M = 131 \cdot 1836 $ a.u. for Xe$^{q+}$ ions, where $E_{k}$ denotes the initial ionic kinetic energy).
For charges $q=35$ and $q=40$, the critical ionic velocities $v_c $ are close to the experimental one (see Table \ref{tab:table1}), indicating that both energies contribute to the crater formation.
The values of $v_c$ for Xe$^{25+}$ and for Xe$^{30+}$ are much smaller than the experimental ones, so that the main contribution in the nanostructure formation gives the deposited kinetic energy $E_{k,dep}$. Concerning the type (shape) of the nanostructures,  the appearance of the nanocraters in experiment is in accord with the prediction of the micro-staircase model. On the other hand, the hillocks have been obtained in experiment with  Xe$^{35+}$  ions \cite{Stabrawa2017} impinging upon the surface of the 50 nm gold layer at velocity 0.19 a.u., while the critical one is 0.22 a.u. \cite{Majkic2021}. In the case of 25 nm titanium nanolayers the experimental velocities for $q= 20, 25, 30$ and 35 were 0.144, 0.16, 0.176 and 0.19, in a.u., respectively. The corresponding  critical velocities are 0.06, 0.16, 0.22 and 0.24, in a.u. \cite{Majkic2021}. The results of the present experiment  and the results of the previous ones \cite{Stabrawa2017, Stabrawa2020, Pomeroy2007} confirm a common conclusion: for the ionic velocities  $v < v_c$ or $v \approx v_c$ the surface modification leads to the nanohillocks formation \cite{Stabrawa2017,Majkic2021}, while for $v > v_c$  the predominant surface structures are the craters (rings) \cite{Stabrawa2020, Pomeroy2007, Majkic2021}.

The neutralization energy $W^{(q,nMV)}$ and the deposited kinetic energy $E_{k,dep}$ can be also connected to the size of the formed nanostructures. The experimental results for the crater diameters show the significant increasing from $q=25$ to $q=40$, see  \ref{fig:Gold_potential_energy}.  For $v_{exp}=0.29$ a.u. and Xe$^{25+}$ ion the diameter $D=12$ nm (226.8 a.u.) and for Xe$^{40+}$ ion  diameter  $D =$ 23.4 nm (442.3 a.u.). The ionic neutralization energy (and also the initial ionic potential energy) exhibits the same increasing behavior ($W^{(25,nMV)}=47$ a.u.  and $W^{(40,nMV)}=162$ a.u.), see Table \ref{tab:table2}.
 \begin{table} [b]
	\begin{center}
		\caption{\label{tab:table2}
			Neutralization energy $W^{(q,nMV)}$, deposited kinetic energy  $E_{k,dep}$ and crater diameter $D$  in the case of the  surface nanocrater formation for $v=v_{exp}=0.29$ a.u.
			in the nMV-system by the impact of Xe$^{q+}$ ions, for   $\Delta x  \approx 38.5$ a.u. }
		\begin{tabular}{cccccc}
			\hline
			\multicolumn{6}{c}{100 nm Au nanolayer}   \\
			\hline
			$q$     &25  &30   &35  &40   & \\
			\hline
			$W^{(q,nMV)}$ (a.u.)   &47   &88   &127  &162 & \\
			\hline
			$E_{k,dep}$ (a.u.)    &557   &569   &577  &587 & \\
			\hline
           $D$ (nm)    &12   &13   &15  &23.4 & \\
			\hline
		\end{tabular}
	\end{center}
\end{table}

The $q$ dependence of the deposited kinetic energy, obtained on the base of Eq. \ref{V}, is less pronounced  ($E_{k,dep}$ for Xe$^{25+}$ is 557 a.u. and for Xe$^{40+}$, $E_{k,dep}= 587$ a.u.), see Table \ref{tab:table2}. For these reasons, it is convenient to present the experimentally obtained crater diameters as a function of the potential (or the neutralization) energy.
	
The velocity effect on the crater diameter $D$ for Xe$^{35+}$ ions we study for the experimental values $v_{exp}= 0.29, 0.33, 0.36 $ and 0.38 a.u. From the experimental results  one  recognize the  weak  decreasing of the quantity $D$  with  increasing of the ionic velocity (kinetic energy) see Fig. \ref{fig:Gold_kinetic_energy}; (for $v= $0.29 a.u. diameter $D= 15$ nm (283.5 a.u.) and for $v= $0.38 a.u. diameter $D = 12$ nm (226.8 a.u.)).
On the other hand, the deposited kinetic energies $ E_{k,dep} $ increase slightly with increasing of $v$ (for $v$=0.29 a.u. $ E_{k,dep} $= 577 a.u. and for $v= $0.38 a.u. $ E_{k,dep}= $635 a.u.), while the neutralization energy $W^{(35,nMV)}$ show a noticeable decreasing character (for $v=$0.29 a.u. $W^{(35,nMV)}=$127 a.u. and for $v= $0.38 a.u. $W^{(35,nMV)}= $35.5 a.u.), see Tab. \ref{tab:table3}.
\begin{table}
	\begin{center}
		\caption{\label{tab:table3}
			Neutralization energy $W^{(q,nMV)}$, deposited kinetic energy  $E_{k,dep}$ and crater diameter $D$  in the case of the  surface nanocrater formation for $v_{exp}=0.29,0.33,0.36$ and 0.38 a.u. in the nMV-system by the impact of Xe$^{35+}$ ions, for  $\Delta x  \approx 38.5$ a.u. }
		\begin{tabular}{cccccccc}
			\hline
			\multicolumn{6}{c}{100 nm Au nanolayer}  \\
			\hline
			$v_{exp}$ (a.u.)   & 0.29 & 0.33  & 0.36   & 0.38  & \\
			\hline
			$W^{(q,nMV)}$ (a.u.)   & 127   & 66.8   & 47.2   & 35.5  & \\
			\hline
			$E_{k,dep}$ (a.u.)  & 577   & 600   & 608   & 612  & \\
			\hline
         $D$ (nm)    &15   &12.9   &13.9  &12.15 & \\
			\hline
		\end{tabular}
	\end{center}
\end{table}

The  role of the neutralization energy $W^{(q,nMV)}$ and the deposited kinetic energy  $E_{k,dep}$  can be more precisely discussed from the relation between the crater diameter and the total deposited energy:
$E_{tot,dep}=E_{k,dep}+ W^{(q,nMV)}$. Assuming that the energy $E_{tot,dep}$ is localized in the cylindrical region  od the diameter $D$ and depth $\Delta x$, we get the relation:
\begin{equation}\label{R}
D=f\sqrt{{E_{k,dep}+W^{(q,nMV)}}},
\end{equation}
where the factor $f$ reflects the  target properties. 
Both energies $E_{k,dep}$ and $W^{(q,nMV)}$ are charge dependent, so that the nanocrater size will express the same behavior.
From Eq. \ref{R} and Tables \ref{tab:table2} and \ref{tab:table3} one conclude that the main  contribution to the diameter $D$ gives the deposited kinetic energy. However, the neutralization energy term  in Eq. \ref{R} must be taken into account in order to obtain the experimentally observed behavior of the crater diameter $D$ discussed in Tables Tab. \ref{tab:table2} and Tab. \ref{tab:table3}: pronounced  increasing od $D$ with increasing of $q$ and weak decreasing of $D$ with increasing of the ionic velocity $v$.  
The discussed significance of the deposited kinetic energy and the role of the neutralization energy  is characteristic for the moderate velocity case used in the  experiment. We note that for the very low velocities, the neutralization energy (close to the potential energy) plays a dominant role. The  increasing of $D$ with increasing of $q$  and the $v$-dependence of the crater diameter  obtained in the present experiment  is in a qualitative agreement  with the prediction of the proposed model.

Within the framework of micro-staircase model, the mechanism of the nanocraters and nanohillocks formation at metallic surfaces is different.  At velocities $v <v_c$ characteristic for the hillock formation, a dominant role has the neutralization process: the strength of the bonds between atoms decreases inducing their stretching. The rearrangement of atoms leads to the rise of the volume above the surface and hillock formation. The deposited energy is insufficient for melting the material (the hillocks are formed without melting). The predicted mechanism of the nanohillock formation on metal surface \cite{Majkic2021} is different in comparison to the thermal spike model used in the case of nanohillock formation on insulator \cite{El-Said2007b}. In the case of crater creation (for $v > v_c$) considered in the present paper, the neutralization (above the surface) induces the lattice vibration and the first destabilization of the target. Inside the solid, the elastic collisions of the charged projectile with target atoms and produced recoils lead to the disordering of the target atoms generating the highly disturbed near surface area $V$.  A large amount of kinetic energy deposits into the solid, resulting in a significant decrease in the target cohesive energy. The strength of the bounds between the target atoms inside the crater volume tends to be zero and a number of atoms are ejected from the surface.  In the intermediate stages of the craters formation, in the centre of the active volume $V$, it is possible that the temperature far exceeds the melting temperature. The deposited neutralization energy during the process above the surface has a small contribution to the nanocrater formation in comparison to the deposited kinetic energy during the collision cascade below the surface; however, the main $q$ and $v$ dependence of the crater size are governed by the neutralization energy.

\subsection{MD simulations}

\begin{figure} [!bt]
	\centering
	\includegraphics[width=8.5cm]{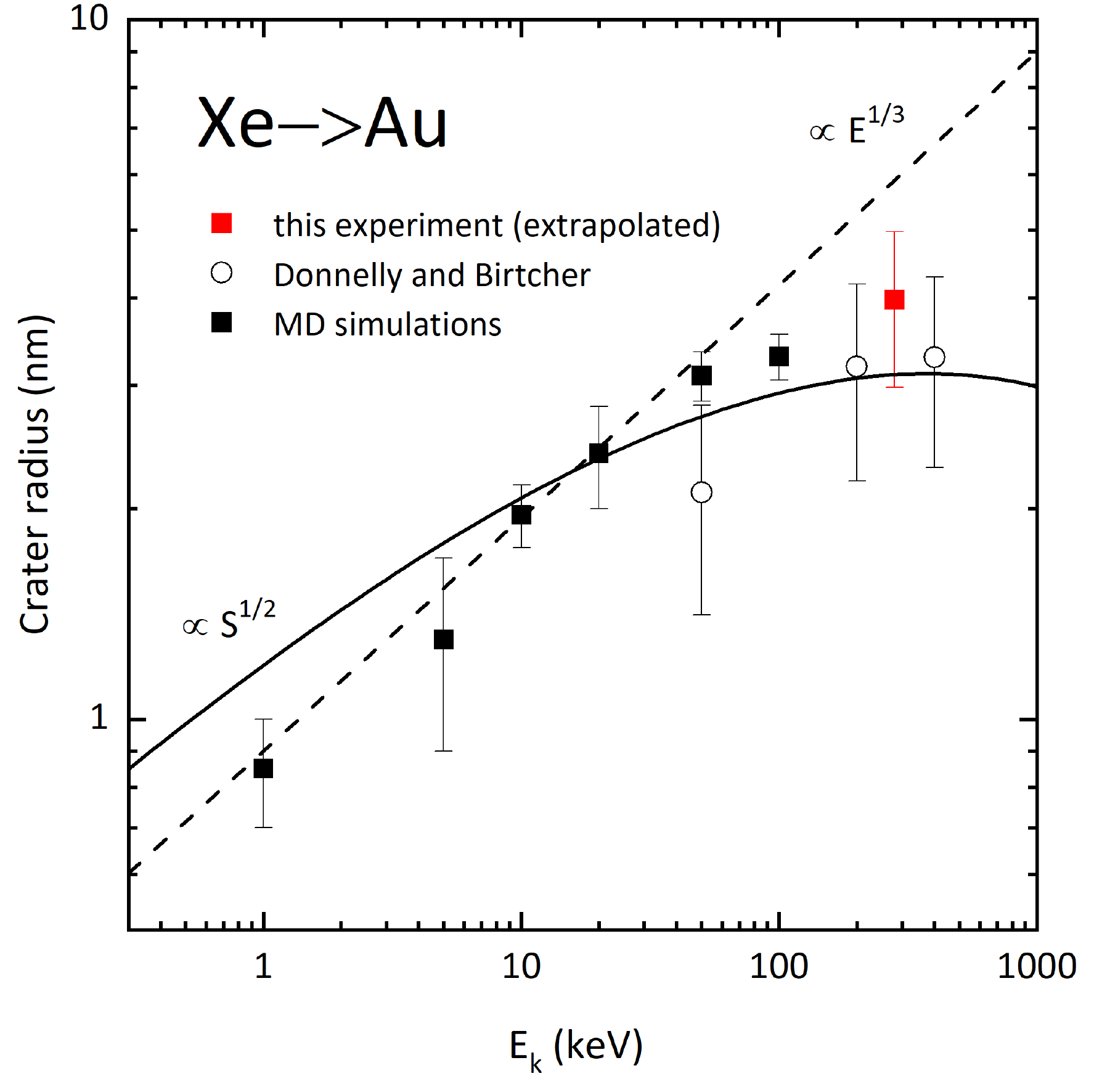}\\
	\caption{Comparison of the results of crater radius, as a function of the ion kinetic energy, obtained by our group with the results of the experiment (Donnelly and Birtcher) for single charged Xe\textsuperscript{+} ion impinging on Au surface \cite{Birtcher1996} and MD simulations \cite{Bringa2001}. In the figure, the nuclear stopping power S$^{1/2}$ (solid line) and ion energy E$^{1/3}$ (dashed line) curves are also presented.
		\label{fig:Gold_single_comparison}
	}
\end{figure}

We  also compare the present experimental results with molecular dynamics (MD) simulations presented in \cite{Bringa2001}. We  compare our results for HCI xenon ion with Xe\textsuperscript{+} ion after fitting the data on  Fig. \ref{fig:Gold_potential_energy}, and further normalization to the potential energy equal to the potential energy of Xe$^{+}$. The results of the comparison we present in Fig. \ref{fig:Gold_single_comparison}. In the figure, the nuclear stopping power S$^{1/2}$ (solid line) and the ion energy E$^{1/3}$ (dashed line) curves are also presented. We obtain very good agreement with the experimental data of Donnelly and Birtcher \cite{Birtcher1996} and MD simulations \cite{Bringa2001}, which confirms the validity of our experimental procedure.

It is important to mention that, very recently, molecular dynamics methodology coupled with two-temperature model (2T-MD) \cite{Duffy2007}, was used by Khara et al. to simulate the structural evolution of bcc metals (Fe and W) and fcc metals (Cu and Ni) following irradiation by SHI (electronic stopping power regime) \cite{Khara2017}. They found that number of material parameters (melting temperature, electronic thermal conductivity and electron-phonon coupling strength), and their electronic properties temperature dependence, have a strong influence on the resistance of metals to damage induced by SHI irradiation. They also showed that high thermal conductivity and relatively low electron-phonon coupling of fcc metals render them relatively insensitive to damage, in spite of their relatively low melting temperatures. The strong electron-phonon coupling of the bcc metals (Fe and W) is primarily responsible for the sensitivity of these metals to damage \cite{Khara2017}. The cited calculations are in contradiction with the experimental  results for Au (fcc metal) - HCI systems, for which we obtain the surface nanocraters in the velocity range $v\in[0.29, 0.38]$ a.u. and the nanohillocks for lower ionic velocities $v\in[0.144, 0.19]$ a.u.  \cite{Stabrawa2017}. 
In the case of nanocrater formation, both the deposited kinetic energy and the neutralization energy participate in  the process; the nanohillocks are formed predominantly by the participation of the neutralization energy.
The  calculations \cite{Khara2017} showed a significantly different response of bcc and fcc metals to the deposition of energy in the interaction of SHI ions with surfaces and encouraged us to undertake such tests for HCI.  At the moment, similar calculations does not exist for HCI, where it is necessary to take into account the neutralization process. The  model proposed here represents a theoretical approach of that kind, stimulated by the experimental findings.

\section{Conclusions}
Understanding of mechanism of the nanostructures creation on metallic surfaces is very important both from the theoretical and possible application point of view. In this paper we have studied Au nanolayers surfaces irradiated by slow highly charged Xe$^{q+}$ ions ($q$ = 25, 30, 35, 36 and 40). For the first time, for such systems,  well pronounced modifications of the nanolayers surfaces, due to impact of the HCI ions, in the form of nanocraters have been observed. This allowed for systematical study of dependence of the size of nanostructures on potential and kinetic energy of the ions. Analysis of the crater diameter $D$ for different initial charge states $q$ of the Xe ions showed  a significant  dependence of the quantity $q$ (expressed via potential energy in Fig. \ref{fig:Gold_potential_energy}). Additionally, for interaction of the Xe$^{35+}$ ions with Au nanolayers the dependence of the structure formation on the ion kinetic energy (280 keV, 360 keV, 420 keV and 480 keV) was studied. Week alteration of the crater diameter (Fig. \ref{fig:Gold_kinetic_energy}) with  the ion kinetic energy was observed in the analyzed energy range. Our results were qualitatively interpreted within the micro-staircase model for the neutralization energy combined by the charge dependent kinetic energy deposition. The experimental results are also compared with the available simulations and the previous experimental data. The results will be potentially of great importance for further development of modern technologies (e.g. single HCI nano-pattering \cite{Gierak2014}, role of the HCI impurities in tokamak plasma-metallic wall interaction \cite{Winter2007}) and will open up many application possibilities (e.g. DNA sequencing or water desalination \cite{Kozubek2019}).

\section*{Acknowledgments}
The equipment was purchased thanks to the financial support of the European Regional Development Fund in the framework of the Polish Innovative Economy Operational Program (contract no. WNP-POIG.02.02.00-26-023/08), the Development of Eastern Poland Program (contract no. POPW .01.01.00-26-013/09-04) and Polish Ministry of Education and Science (project 28/ 489259/SPUB/SP/2021). N. N. Nedeljkovi\'{c} and M. D. Majki\'{c} are grateful for the support of the Ministry of Education, Science and Technological Development of the Republic of Serbia (projects 171016, 171029).

\bibliography{stabrawa2022_revised_2}

\end{document}